\documentclass[12pt]{article}
\usepackage{graphicx}

\def\baselinestretch{1.2}
\newcommand{\be}{\begin{equation}}
\newcommand{\ee}{\end{equation}}
\newcommand{\beq}{\begin{eqnarray}}
\newcommand{\eeq}{\end{eqnarray}}

\begin{document}
\begin{titlepage}

\begin{flushright}
hep-th/0304181\\
\end{flushright}
%\vspace{12 mm}

\vfil\vfil

\begin{center}

{\Large{\bf Black holes in G\"{o}del universes and pp-waves}}

\vfil

\vspace{5mm}

Eric G. Gimon and Akikazu Hashimoto\\

\vspace{10mm}

Institute for  Advanced Study\\ School of Natural Sciences\\
Einstein Drive, Princeton, NJ 08540\\

\vfil

\end{center}

\begin{abstract}
\noindent    
We find exact rotating and non-rotating neutral black hole solutions
in the G\"{o}del universe of the five dimensional minimal supergravity
theory.  We also describe the embedding of this solution in M-theory.
After dimensional reduction and T-duality, we obtain a supergravity
solution corresponding to placing a black string in a pp-wave
background.
\end{abstract}

\vspace{0.5in}

\end{titlepage}
\renewcommand{\baselinestretch}{1.05}  %Line spacing

The G\"{o}del universe \cite{Godel:1949ga} is an exact solution of
Einstein's equations in the presence of a cosmological constant and
homogeneous pressureless matter.  This space-time solution exhibits
several peculiar features including in particular the presence of
closed time-like curve through every point. Recently, a space-time
exhibiting most of the peculiar features of G\"{o}del universes was
shown to be an exact solution of minimal supergravity in 4+1
dimensions, preserving some number of supersymmetries
\cite{Gauntlett:2002nw,Herdeiro:2002ft}.  As a result, these solutions
can be embedded in supergravity theories of 10 or 11 dimensions and
may constitute consistent backgrounds of string theory.  The
consistency of this solution was further investigated recently by
\cite{Boyda:2002ba,Harmark:2003ud} who found that the supersymmetric
G\"{o}del universes of \cite{Gauntlett:2002nw,Herdeiro:2002ft} are
related by T-duality to the pp-wave solutions, which have generated
significant interest recently following the works of
\cite{Blau:2001ne,Berenstein:2002jq}.

In this article, we construct a space-time describing a
Schwarzschild black hole localized inside the G\"{o}del universe,
and describe some of its basic properties. Just as for Minkowski,
anti de Sitter, and de Sitter spaces, such a solution provides
important insights into the nature of G\"{o}del universes. We
should mention that a different but very interesting space-time
describing a charged extremal black hole with finite horizon area
in a G\"{o}del universe was identified in \cite{Herdeiro:2002ft}.

Finding space-times describing a black object localized inside the
pp-wave is an important outstanding problem.  Discussions of the
recent approaches to this problem can be found in
\cite{Hubeny:2002pj,Hubeny:2002nq,Liu:2003ct}.  As a bonus for
finding the black hole solution in G\"{o}del universes, we are
able to construct the black string solution in pp-waves.

We will follow \cite{Gauntlett:2002nw,Herdeiro:2002ft} and work mostly
with the bosonic components of the 4+1 dimensional minimal
supergravity.  Unlike \cite{Gauntlett:2002nw,Herdeiro:2002ft},
however, we will not require our solutions to preserve any
supersymmetry. The bosonic fields of the minimal supergravity theory
in 4+1 dimensions consist of the metric and a 1-form gauge
field. Their equations of motion are given by
\be R_{\mu \nu} = 2 \left(F_{\mu \alpha} F_{\nu}{}^\alpha - {1 \over 6} g_{\mu \nu} F^2 \right) , \qquad
D_\mu F^{\mu \nu} = {1 \over 2 \sqrt{3}} \tilde \epsilon^{\alpha \beta \gamma \mu \nu} F_{\alpha \beta} F_{\gamma \mu} \ , \label{eqm} \ee
where $\tilde \epsilon_{\mu \nu \lambda \kappa \sigma} = \sqrt{-\det
g} \, \epsilon_{\mu \nu \lambda \kappa \sigma}$. We follow mostly the
conventions of \cite{Gauntlett:2002nw,Herdeiro:2002ft} and use angular
coordinates
\be
x^1 + i x^2 = r \cos {\theta \over 2} e^{i\left({\psi + \phi \over2}\right)}, \qquad
x^3 + i x^4 = r \sin {\theta \over 2} e^{i\left({\psi - \phi \over2}\right)} \ . \ee
The G\"{o}del universe is a solution of the equations of motion
(\ref{eqm}) given by
\beq ds^2 &=& -(dt + j r^2 \sigma^3_L)^2 + dr^2 + {r^2 \over 4}
\left( d \theta^2 + d \psi^2 + d \phi^2 + 2 \cos\theta\, d \psi d
\phi\right), \cr A &=& {\sqrt{3} \over 2} j r^2
\sigma^3_L \eeq
where,  following the conventions of \cite{Gauntlett:2002nw,Herdeiro:2002ft},
\be \sigma_L^3  =  d \phi + \cos \theta \, d \psi \ . \ee
The parameter $j$ sets the scale of the background, which reduces to
Minkowski space for small $j$.  From the sign of the $g_{\phi\phi}$
component of the metric, it is easy to recognize a closed time-like
curve parameterized by $\phi$ with all other coordinates fixed, for $r
> 1/2j$. We will refer to the surface at fixed $r$ where $g_{\phi
\phi}$ vanishes as the ``velocity of light surface.'' It should be
emphasized, however, that since the G\"{o}del space-time is
homogeneous, there is a closed time-like curve going through every
point in space-time.

In order to find a solution to the supergravity equations of motion
describing black holes in this background, consider an ansatz
 \beq
ds^2 &=& -f(r)\, dt^2 - g(r)\,  r \sigma^3_L dt - h(r)\, r^2
(\sigma^3_L)^2 \cr && + k(r)\,   dr^2 + {r^2 \over 4} \left( d
\theta^2 + d \psi^2 + d \phi^2 + 2 \cos\theta\, d \psi d
\phi\right), \cr 
A &=& {\sqrt{3} \over 2} j r^2 \sigma^3_L \ .
\label{ansatz}
 \eeq
Substituting this ansatz into the equations of motion gives rise
to a rather complicated set of equations. It can be easily
checked, however that the choice
\beq
f(r) & = & 1 - {2m \over r^2} \cr
g(r) & = & 2 j r \cr
h(r) & = & j^2  \left(r^2 + 2 m\right) \cr
k(r) & = & \left(1 - {2m \over r^2} + {16 j^2 m^2 \over r^2}\right)^{-1}   
\label{godelbh}
\eeq
solves these equations.  In the small $j$ limit, this solution reduces
to an ordinary Schwarzschild black hole in 4+1 dimensions.  On the
other hand, in the small $m$ limit, we recover the G\"{o}del universe.
We therefore conclude that this background corresponds to a
Schwarzschild black hole placed inside the G\"{o}del universe. This
solution is the main result of this paper; we will study its various
properties.

The ansatz (\ref{ansatz}) clearly preserves 5 of the 9 isometries
of the G\"{o}del universe. They are generated by time translation
$\partial_t$, as well as the $SU(2) \times U(1)$ subgroup of the
$SO(4) = SU(2) \times SU(2)$ isometry group on $S^3$
\be \parbox{5in}{
\beq
\xi_1^R & = & -\cot \theta \cos \psi \, \partial_\psi - \sin \psi \, \partial_\theta + {\cos \psi \over \sin \theta} \, \partial_\phi \cr
\xi_2^R & = & - \cot \theta \sin \psi \, \partial_\psi + \cos \psi \, \partial_\theta + {\sin\psi \over \sin \theta} \,  \partial_\phi \cr
\xi_3^R & = & \partial_\psi , \cr 
\xi_3^L & = & \partial_\phi \  . \nonumber % \label{isometries} 
\eeq
}   \label{isometries} \ee

The most salient feature of the solution (\ref{godelbh}) is the
existence of a horizon at
\be r_{BH}^2 =  2m(1 - 8 j^2 m) \ . \ee
This surface is a horizon in the sense that future directed lightcones
emanating from all points inside the horizon are strictly contained
inside.  The surface $r = r_{BH}$ also exhibits many of the standard
properties of a black hole horizon, as we will see below.

The area of the horizon is
\be \mathcal{A} =2 \pi^2 \sqrt{8 m^3(1 - 8 j^2 m)^5} \ . \label{areabh}\ee
It is tempting to apply the standard interpretation of black hole
thermodynamics and think of
\be S = {\mathcal{A} \over 4 G_5} \ee
as the entropy.  It is interesting to note that with this definition
the entropy does not increase monotonically with $m$.  In fact, at $m
= 1/8 j^2$, the horizon area vanishes. For $m$ greater than this
critical value, the horizon disappears and the space-time
(\ref{godelbh}) is nakedly singular. This appears to constitute a
physical upper bound on a mass of a neutral state in a G\"odel
universe, rendering the number of degrees of freedom to be finite.

As one increases $m$, the velocity of light surface also moves in
toward small $r$
\be r^2_v = {1 \over 4 j^2} (1 - 8 j^2 m)  \ . \ee

One of the main objectives of \cite{Boyda:2002ba} was to study the
holography of G\"{o}del universes from a ``phenomenological'' point of
view and to compare against the known properties of de Sitter
space. To this end, the authors of \cite{Boyda:2002ba} computed the
location of a ``preferred holographic screen'' for a timelike observer
at the origin, as defined in \cite{Bousso:1999cb}. Due to the $SU(2)$
symmetry (\ref{isometries}), this simply amounts to computing the
maximum area for a family of surfaces with 3-sphere topology, each at
fixed $t$ and $r$ coordinates.  For the space-time (\ref{godelbh}),
each surface has area
\be \mathcal{A}(r) = r^3 \sqrt{1 - 4j^2(r^2 + 2m)} \ . \ee
For $m=0$, this expression reduces to the result obtained in
\cite{Boyda:2002ba}. The location and area of the preferred
holographic screen come out to be
\be r_s^2    = {3(1 - 8 j^2 m) \over 16 j^2} \ , \qquad \mathcal{A}_s = {3 \sqrt{3}\pi^2  (1 - 8 j^2 m)^2 \over 64 j^3} \ . \label{areaps} \ee
Note that the area $\mathcal{A}_s$ of the preferred screen decreases
as we increase $m$.  Also, at a finite value of $m=3/32j^2$, but
before the critical value for the appearance of a naked singularity $m
= 1/8j^2$, the black hole horizon and the preferred screen coincide.
This convergence is similar to what was observed in de Sitter space
\cite{Gibbons:1977mu}, except that in that case screen and horizon
meet just as a naked singularity appears.

The curvature invariants for this space-time take a simple form. For
example, the Ricci scalar is
\be R = {16 j^2(r^2 - m) \over r^2} \ . \ee
The space-time is smooth at the horizon radius as long as $m < 1/8 j^2$,
but there is always a singularity at $r=0$ for non-vanishing $m$.

\begin{figure}
\centerline{\includegraphics[width=2in]{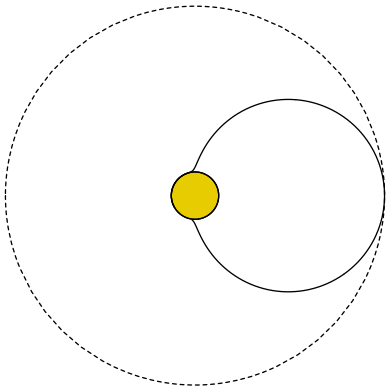}}
\caption{The projection to fixed $(t,\theta,\psi)$ plane of a null
geodesic which is tangent to the velocity of light surface at an
instant and spirals in towards the horizon in the future and in the
past. The dotted line is the velocity of light surface and the shaded
region is the region inside the horizon.
\label{figa}}
\end{figure}

It possible to find an analytic form for geodesics in this space-time
geometry following the approach of \cite{Boyda:2002ba}.  Let us write
the tangent vector to the geodesic as
\be \xi =
\dot t {\partial \over \partial t}  +
\dot r {\partial \over \partial r} +
\dot \theta {\partial \over \partial \theta} +
\dot \psi {\partial \over \partial \psi} +
\dot \phi {\partial \over \partial \phi} \ee
where the dot denotes derivative $d/d \lambda$ with respect to the
affine parameter $\lambda$, and define the integrals of motion
\be (\xi,\xi) = -M^2, \qquad (\xi,\partial_t) = -E , \qquad (\xi,\partial_\psi) = L_\psi, \qquad  (\xi,\partial_\phi) = L_\phi \ . \ee
It is useful to define one more integral of motion
\be \Xi = (\xi,\xi_1^R) = \cos\psi \, \left(L_\psi \cos\theta - {L_\phi \over \sin\theta}  \right) - {r^2 \dot \theta^2 \sin(\phi) \over 4} \ , \ee
where $\xi^R_1$ is one of the isometries listed in
(\ref{isometries}). The constraints implied by these conserved
quantities simplify drastically if we set $L_{\phi} = L_\phi = \Xi
= 0$ and read
\be \dot t = {(r^2(1 - 4 j^2 (r^2 + 2m)) \over r^2 - 2m + 16 j^2 m^2},   \qquad  \dot \theta = 0 , \qquad \dot\psi = 0 , \qquad \dot\phi = {4 j r^2 \over  r^2 - 2m + 16 j^2 m^2}\ . \label{constraints} \ee
If we set $M=0$ so as to consider only null geodesics, we find a very
simple expression for the radial derivative
\be \dot r^2 =   ( (1 - 8 j^2 m) -  4  j^2 r^2   ) \ee
which is easily solved. Integrating (\ref{constraints}) then gives
the full expression for the geodesics in this family. If we pick
initial conditions so that
\be r(\lambda) = {1 \over 2j} \sqrt{1 - 8 j^2 m} \cos(2 j  \lambda) \ , \ee
the corresponding geodesic is tangent to the velocity of light surface
at $\lambda = 0$ and asymptotes to a trajectory
\be r^2 = r_{BH}^2 = 2 m(1 - 8 j^2 m), \qquad \phi = {(1 - 8 j^2 m)^2 \over 4j} t + \phi_0\ , \ee
which spirals into the horizon in infinite coordinate time $t$
(although it crosses the horizon in finite affine time). See
figure \ref{figa} for an illustration. This is the behavior
expected for a geodesic falling towards a black-hole horizon. One
can also compute the expansion scalar for the congruence
associated with this family of geodesics
\be \Theta = {(3 - 8 j^2(2r^2+3m)) \over r \sqrt{1 - 4 j^2 (r^2+2m)}} \ee
which vanishes at the radius of the preferred screen $r = r_s$ as expected.

It is interesting to consider the embedding of the solution
(\ref{godelbh}) of 4+1 dimensional supergravity into string
theory. The most efficient way to do this is to first embed the
solution of supergravity in 4+1 dimensions into 10+1 dimensions.
This can be done by considering an ansatz
\be ds^2 = g_{\mu \nu} dx^\mu dx^\nu + \sum_{i=5}^{10} (dx^i)^2, \qquad C = {2 \over \sqrt{3}} A \wedge K \ , \label{11dansatz} \ee
where
\be K = dx^5 \wedge dx^6 + dx^7 \wedge dx^8 + dx^9 \wedge dx^{10} \ . \ee
Indices $\mu$ and $\nu$ run from 0 to 4.  One can verify that
substituting this ansatz into the supergravity equations of motion in
10+1 dimensions
\be R_{ab} - {1 \over 12} \left( F_{acde} F_{b}{}^{cde} - {1 \over 12} g_{ab} F^2\right) \ee
gives rise precisely to the equations of motion for supergravity in 4+1
dimensions (\ref{eqm}). One can now dimensionally reduce along the
$x^{10}$ coordinate to obtain the solution (\ref{godelbh}) embedded
into type IIA supergravity.

In the absence of the black hole, T-dualizing along $x^9$ will
give rise to a pp-wave solution obtained by taking the Penrose
limit of intersecting D3-branes. In the presence of the black
hole, T-duality yields a new black string solution for the type
IIB theory
\beq ds^2 &=& -\left(1 - {2 m \over r^2} \right) dt^2 +dy^2 - 2 j r^2 \sigma^3_L (dt-dy)  -  2 m j^2 r^2 (\sigma^3_L)^2 \label{blackstring}\\
&&  + \left(1 - {2 m \over r^2} + {16 j^2 m^2 \over r^2} \right)^{-1} dr^2 + {r^2 \over 4} \left( d \theta^2 + d \psi^2 + d \phi^2 + 2 \cos\theta\, d \psi d \phi\right) + ds^2_{T^4} \nonumber
\eeq
where $y$ is the coordinate dual to $x^9$.  There are also non-trivial
NS-NS and R-R field strengths in the background which can be found by
following the duality starting from (\ref{11dansatz}).

In the limit $m \rightarrow 0$, this new solution
(\ref{blackstring}) reduces to a well known pp-wave geometry. In
the $j \rightarrow 0$ limit, the solution reduces to uncharged
black string solution of type IIB supergravity. This suggests that
(\ref{blackstring}) should be interpreted as the solution
describing a black string in an asymptotically pp-wave background
geometry.

There is a subtlety with this interpretation, however. If one collects
the angular part of the metric (\ref{blackstring}) by fixing $t$, $y$, and $r$, one finds
\be ds^2 = {r^2 \over 4} \left( (\sigma_L^1)^2 + (\sigma_L^2)^2 + (1 - 8 j^2 m)  (\sigma_L^3)^2 \right) \ . \ee
In other words, the term in the deformed metric along $\sigma_L^3$ is
of the same order in $r$ as the undeformed metric of the round
3-sphere.  Turning on $m$ therefore has the effect of squashing this
3-sphere until one reaches $m = 1/8j^2$ at which point the 3-sphere is
squashed completely.  Because of this effect on the asymptotic
geometry, the solution (\ref{blackstring}) cannot be interpreted as
that of a black string with the same asymptotic background geometry as
the empty pp-wave.

An important question is whether or not it is possible to find a
solution which would describe a black string which does not squash or
otherwise affect the large $r$ asymptotic geometry of the pp-wave. It
may be that a black hole, as opposed to a black string, will have
smaller effect on this asymptotic geometry.  Similar gravitational
back reaction effects due to localized and delocalized extremal
sources were observed in \cite{Lunin:2002iz}. Perhaps finite energy
density uniformly distributed along the light-cone coordinates of the
pp-wave generically back reacts to deform the large $r$ asymptotics.
It would also be interesting to examine how the Laflamme-Gregory
instability of the black string solution \cite{Gregory:1993vy} is
affected when $j$ is non-vanishing.

It would also be interesting to explore various generalizations of the
solution (\ref{godelbh}). For example, one can easily verify that
\beq
f(r) & = & 1 - {2m \over r^2} \cr
g(r) & = & 2 j r + {2 m l \over r^3}  \cr
h(r) & = & j^2  \left(r^2 + 2 m\right) - {m l^2 \over 2 r^4} \cr
k(r) & = & \left(1 - {2m \over r^2} + {16 j^2 m^2 \over r^2} + {8 j m l \over r^2} + {2 m l^2 \over r^4}\right)^{-1} \label{godelspinningbh}
\eeq
also solves the equations of motion. By taking the
$j\rightarrow 0$ limit, we recover the rotating black hole
solution of 4+1 gravity \cite{Myers:1986un}.  One could therefore
think of this solution as a rotating black hole inside the
G\"{o}del universe. It is possible, in particular, to tune the
angular momentum
\be l = -4 j m \ee
so as to make the horizon angular velocity $\Omega_H$ vanish.

It should also be possible to find a generalization to charged black
holes by following the construction outlined in
\cite{Breckenridge:1996sn,Cvetic:1996xz,Horowitz:1996ay}. This will
give rise to the non-extremal generalization of the black hole solution
identified in \cite{Herdeiro:2002ft}.

There are many other interesting issues to explore. One would like
to map out the full causal structure of the black hole solution
(\ref{godelbh}).  It would also be interesting to find a suitable
generalization of ADM mass and angular momentum to the G\"{o}del
universe and to compute their values for (\ref{godelbh}) and
(\ref{godelspinningbh}). Ultimately, one would like to understand
the physical meaning of the area of the preferred screen
(\ref{areaps}) and the horizon (\ref{areabh}) as being related to
some microscopic state counting, possibly of strings in a pp-wave
background. We hope that the explicit solution of supergravity
equations of motion describing black holes in this background will
stimulate further insight into these fascinating issues.

\section*{Acknowledgements}

We would like to thank
S.~Cherkis,
M.~Cvetic,
G.~Horowitz,
N.~Itzhaki,
and
J.~Maldacena
for discussions. The work of AH is supported in part by DOE grant
DE-FG02-90ER40542 and the Marvin L.~Goldberger Fund.  EGG is
supported by NSF grant PHY-0070928 and by Frank and Peggy Taplin.

\bibliography{godel}\bibliographystyle{utphys}

\providecommand{\href}[2]{#2}\begingroup\raggedright\begin{thebibliography}{10}

\bibitem{Godel:1949ga}
K.~G{\oo}del, ``An example of a new type of cosmological solutions of
  Einstein's field equations of graviation,'' {\em Rev. Mod. Phys.} {\bf 21}
  (1949)
447--450.
%%CITATION = RMPHA,21,447;%%.

\bibitem{Gauntlett:2002nw}
J.~P. Gauntlett, J.~B. Gutowski, C.~M. Hull, S.~Pakis, and H.~S. Reall, ``All
  supersymmetric solutions of minimal supergravity in five dimensions,''
\href{http://www.arXiv.org/abs/hep-th/0209114}{{\tt hep-th/0209114}}.
%%CITATION = HEP-TH 0209114;%%.

\bibitem{Herdeiro:2002ft}
C.~A.~R. Herdeiro, ``Spinning deformations of the D1-D5 system and a geometric
  resolution of closed timelike curves,''
\href{http://www.arXiv.org/abs/hep-th/0212002}{{\tt hep-th/0212002}}.
%%CITATION = HEP-TH 0212002;%%.

\bibitem{Boyda:2002ba}
E.~K. Boyda, S.~Ganguli, P.~Ho\v{r}ava, and U.~Varadarajan, ``Holographic
  protection of chronology in universes of the G{\oo}del type,''
\href{http://www.arXiv.org/abs/hep-th/0212087}{{\tt hep-th/0212087}}.
%%CITATION = HEP-TH 0212087;%%.

\bibitem{Harmark:2003ud}
T.~Harmark and T.~Takayanagi, ``Supersymmetric G{\oo}del universes in string
  theory,''
\href{http://www.arXiv.org/abs/hep-th/0301206}{{\tt hep-th/0301206}}.
%%CITATION = HEP-TH 0301206;%%.

\bibitem{Blau:2001ne}
M.~Blau, J.~Figueroa-O'Farrill, C.~Hull, and G.~Papadopoulos, ``A new maximally
  supersymmetric background of IIB superstring theory,'' {\em JHEP} {\bf 01}
  (2002) 047,
\href{http://www.arXiv.org/abs/hep-th/0110242}{{\tt hep-th/0110242}}.
%%CITATION = HEP-TH 0110242;%%.

\bibitem{Berenstein:2002jq}
D.~Berenstein, J.~M. Maldacena, and H.~Nastase, ``Strings in flat space and pp
  waves from N = 4 super Yang Mills,'' {\em JHEP} {\bf 04} (2002) 013,
\href{http://www.arXiv.org/abs/hep-th/0202021}{{\tt hep-th/0202021}}.
%%CITATION = HEP-TH 0202021;%%.

\bibitem{Hubeny:2002pj}
V.~E. Hubeny and M.~Rangamani, ``No horizons in pp-waves,'' {\em JHEP} {\bf 11}
  (2002) 021,
\href{http://www.arXiv.org/abs/hep-th/0210234}{{\tt hep-th/0210234}}.
%%CITATION = HEP-TH 0210234;%%.

\bibitem{Hubeny:2002nq}
V.~E. Hubeny and M.~Rangamani, ``Generating asymptotically plane wave
  spacetimes,'' {\em JHEP} {\bf 01} (2003) 031,
\href{http://www.arXiv.org/abs/hep-th/0211206}{{\tt hep-th/0211206}}.
%%CITATION = HEP-TH 0211206;%%.

\bibitem{Liu:2003ct}
J.~T. Liu, L.~A. Pando~Zayas, and D.~Vaman, ``On horizons and plane waves,''
\href{http://www.arXiv.org/abs/hep-th/0301187}{{\tt hep-th/0301187}}.
%%CITATION = HEP-TH 0301187;%%.

\bibitem{Bousso:1999cb}
R.~Bousso, ``Holography in general space-times,'' {\em JHEP} {\bf 06} (1999)
  028,
\href{http://www.arXiv.org/abs/hep-th/9906022}{{\tt hep-th/9906022}}.
%%CITATION = HEP-TH 9906022;%%.

\bibitem{Gibbons:1977mu}
G.~W. Gibbons and S.~W. Hawking, ``Cosmological event horizons, thermodynamics,
  and particle creation,'' {\em Phys. Rev.} {\bf D15} (1977)
2738--2751.
%%CITATION = PHRVA,D15,2738;%%.

\bibitem{Lunin:2002iz}
O.~Lunin, J.~Maldacena, and L.~Maoz, ``Gravity solutions for the D1-D5 system
  with angular momentum,''
\href{http://www.arXiv.org/abs/hep-th/0212210}{{\tt hep-th/0212210}}.
%%CITATION = HEP-TH 0212210;%%.

\bibitem{Gregory:1993vy}
R.~Gregory and R.~Laflamme, ``Black strings and p-branes are unstable,'' {\em
  Phys. Rev. Lett.} {\bf 70} (1993) 2837--2840,
\href{http://www.arXiv.org/abs/hep-th/9301052}{{\tt hep-th/9301052}}.
%%CITATION = HEP-TH 9301052;%%.

\bibitem{Myers:1986un}
R.~C. Myers and M.~J. Perry, ``Black holes in higher dimensional space-times,''
  {\em Ann. Phys.} {\bf 172} (1986)
304.
%%CITATION = APNYA,172,304;%%.

\bibitem{Breckenridge:1996sn}
J.~C. Breckenridge {\em et al.}, ``Macroscopic and microscopic entropy of
  near-extremal spinning black holes,'' {\em Phys. Lett.} {\bf B381} (1996)
  423--426,
\href{http://www.arXiv.org/abs/hep-th/9603078}{{\tt hep-th/9603078}}.
%%CITATION = HEP-TH 9603078;%%.

\bibitem{Cvetic:1996xz}
M.~Cvetic and D.~Youm, ``General rotating five dimensional black holes of
  toroidally compactified heterotic string,'' {\em Nucl. Phys.} {\bf B476}
  (1996) 118--132,
\href{http://www.arXiv.org/abs/hep-th/9603100}{{\tt hep-th/9603100}}.
%%CITATION = HEP-TH 9603100;%%.

\bibitem{Horowitz:1996ay}
G.~T. Horowitz, J.~M. Maldacena, and A.~Strominger, ``Nonextremal black hole
  microstates and U-duality,'' {\em Phys. Lett.} {\bf B383} (1996) 151--159,
\href{http://www.arXiv.org/abs/hep-th/9603109}{{\tt hep-th/9603109}}.
%%CITATION = HEP-TH 9603109;%%.

\end{thebibliography}\endgroup
\end{document}